# Adaptive on-chip control of nano-optical fields with optoplasmonic vortex nanogates

Svetlana V. Boriskina and Björn M. Reinhard

*Department of Chemistry & The Photonics Center, Boston University, Boston, MA 02215, USA*
*sboriskina@gmail.com, bmr@bu.edu*

**Abstract:** A major challenge for plasmonics as an enabling technology for quantum information processing is the realization of active spatio-temporal control of light on the nanoscale. The use of phase-shaped pulses or beams enforces specific requirements for on-chip integration and imposes strict design limitations. We introduce here an alternative approach, which is based on exploiting the strong sub-wavelength spatial phase modulation in the near-field of resonantly-excited high-Q optical microcavities integrated into plasmonic nanocircuits. Our theoretical analysis reveals the formation of areas of circulating powerflow (optical vortices) in the near-fields of optical microcavities, whose positions and mutual coupling can be controlled by tuning the microcavities parameters and the excitation wavelength. We show that optical powerflow though nanoscale plasmonic structures can be dynamically molded by engineering interactions of microcavity-induced optical vortices with noble-metal nanoparticles. The proposed strategy of re-configuring plasmonic nanocircuits via locally-addressable photonic elements opens the way to develop chip-integrated optoplasmonic switching architectures, which is crucial for implementation of quantum information nanocircuits.

1. Introduction

Plasmonics, which exploits reversible conversion of propagating light into surface charge density oscillations of free electrons in metals – surface plasmons (SPs) – has become a mature technology for nanoimaging and bio(chemical) sensing, and holds high promise for implementation of chip-scale information processing networks [1-3]. Efficient delivering of optical energy into deeply sub-wavelength areas via the excitation of localized SP resonances in metal nanostructures facilitates dramatic enhancements of local field intensities and light-matter interactions. A new area of intense research effort in plasmonics is robust *on-chip* dynamic spatiotemporal manipulation of the sub-wavelength fields and of interactions between single photons and single quantum emitters [4-8]. One important factor limiting the dynamic tunability range of plasmonic nanoelements is the required refractive index modulation range, which – due to shrinking of the spatial light-matter interaction lengths – increases with decreasing spatial dimensions. Another factor is the short dephasing times of the localized SP modes stemming from their high dissipative and radiative losses. The consequences of the latter factor are twofold. First, the reduced 'temporal interaction length' of light with the material imposes even more stringent requirements to the index modulation range. Second, as modes linewidths are inversely proportional to their dephasing times, plasmonic nanostructures typically feature broad scattering spectra consisting of overlapping resonance peaks corresponding to different SP modes. Individual bright and dark SP modes of complex plasmonic nanostructures can be detected by electron-energy-loss spectroscopy [9] and confocal two-photon photoluminescence mapping [10]. However, dynamical selective activation and on/off switching of spatially- and spectrally-distinct modes by using simple controls such as excitation field wavelength or polarization is not effective. On the other hand, the use of temporally phase- and amplitude-modulated pulses and beams [11-16], nanofluidic chambers [17], or elastomeric substrates [18, 19] may complicate on-chip integration of the plasmonic elements. Finally, efficient adaptive focusing of light generated by external sources cannot affect the radiative properties of embedded quantum emitters as it does not change the local density of optical states (LDOS) on the plasmonic chip.

We have recently demonstrated that embedding photonic elements (optical microcavities) into nanoplasmonic circuits introduces a mechanism of strong spectral selectivity into plasmonic networks owing to efficient photon trapping and re-cycling in microcavities in the form of high-Q photonic modes (e.g., whispering gallery (WG) modes) [20, 21]. The resulting optoplasmonic circuits can also perform the functions of *long-range light transfer* with subsequent nanoscale localization as well as spectral and spatial signal (de)multiplexing. Furthermore, embedded high-Q microcavities strongly modify the local density of optical states at specific spatial locations and defined frequencies corresponding to the high-Q cavity modes [20, 22-24], and thus pave the way to tailoring light interactions with quantum emitters. In this paper, we will demonstrate that high-Q photonic elements also provide a rich electromagnetic field phase landscape, which can be used to achieve *on-chip* dynamical light spatial reconfiguring and switching within *nanoscale-size* plasmonic nanostructures, thus completely eliminating the need for bulky external far-field optics.

## 2. Resonant modulation of electromagnetic field phase with optical microcavities

A known but often overlooked feature of optical resonances is strong modification of the *local phase structure* of the wave field. Already the non-resonant light focusing with optical microcavities is accompanied by the emergence of phase anomalies due to the sudden transition from a converging to a diverging wave front [25]. Excitation of the localized resonances causes even more significant modifications of the fine structure of the electromagnetic field phase [26]. They are typically accompanied by a 180$^o$ phase shift of the wave function at the resonance frequency. The effect of the resonant phase reversal can be exploited to engineer Fano-type resonances resulting from the interference between alternative excitation paths of coupled modes, as has recently been demonstrated in a variety of photonic and plasmonic nanostructures [26-29].

The abrupt phase reversal is also a characteristic feature of high-Q photonic resonances in microcavities, such as WG modes in microspheres [30]. This effect is demonstrated in Fig. 1, which shows the calculated far-field scattering and near-field intensity spectra of a polystyrene microsphere illuminated by a y-polarized plane wave (Fig. 1(a,b)). As can be seen in Fig. 1, along with the well-known effect of the resonant field enhancement at the frequency of the WG eigenmode in the microsphere, the resonant reversal of the phase of the major electromagnetic field component is observed. Fig. 1(c) and Media 1 reveal the detailed picture of the evolution of the spatial field intensity distribution in the x-z plane as the excitation frequency is varied and passes through the microcavity WG-mode resonance. The corresponding evolution of the spatial distribution of the phase of the complex-valued principal E-field component $E_y = |E_y| \exp\{\arg(E_y)\}$ is traced in Media 2 (Fig. 1(d)). Clearly, the phase portrait of the WG mode field undergoes a complex spatial transformation, which cannot be easily reduced to a simple coupled-oscillator model [26, 31] frequently invoked to engineer coupled dipole LSP resonances of plasmonic nanoparticles [32]. In particular, the resonant build-up of the optical field intensity corresponding to the formation of the WG-mode pattern is accompanied by a complex picture of nucleation, drift and annihilation of multiple phase singularities both inside the cavity material and in the evanescent field area outside the microsphere (Media 2). The observed phase singularities appear at the points of zero field intensity, which necessarily have an undefined (singular) phase, with the whole 2π range of phase values co-existing at this point [33]. The spatial drift of these phase singularities can be dynamically manipulated by tuning the external control parameters such as the excitation wavelength or the cavity material characteristics [34].

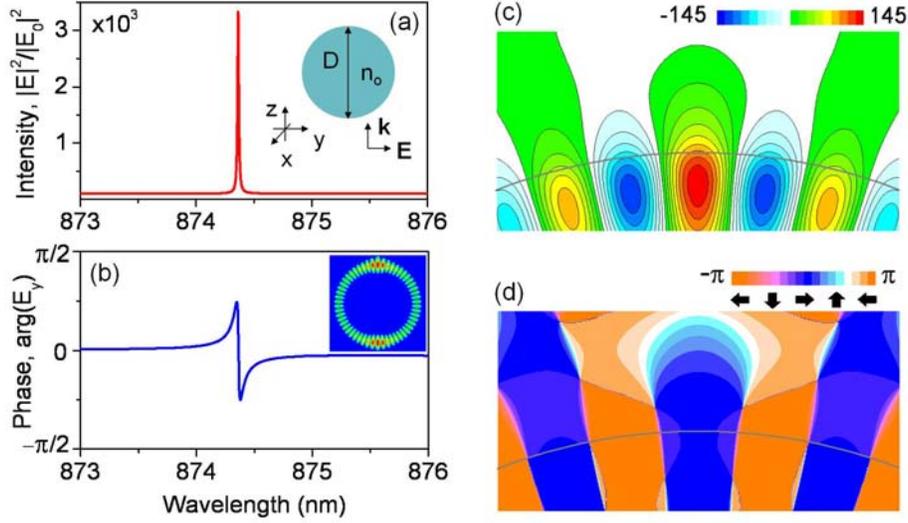

Fig. 1. Evolution of the electric field intensity and phase in the vicinity of a high-Q $TE_{27,1}$ WGM resonance in a single 5.6μm-diameter polysterene microsphere. (a,b) Calculated frequency spectra of the total electric field intensity, $|E|^2$ (a), and a phase of the major E-field component, $Arg(E_y)$ (b). The field is evaluated at the point on the sphere axis 100nm above the surface. The insets show a schematic of the scattering problem geometry (a) and the WGM field intensity map at the resonance. (c,d) Single-frame excerpts from movies of the spatial maps of $Re(E_y)$ (Media 1, c) and $Arg(E_y)$ (Media 2, d) evolution as a function of wavelength (shown at the $TE_{27,1}$ resonance frequency).

The results shown in Fig. 1 (and throughout the rest of the paper) are calculated by using the generalized multi-particle Mie theory (GMT) [20, 35-38]. To demonstrate a general physical concept rather than custom-tailored engineering solutions, we focus on model structures composed of micro- and nano-spherical particles, for which GMT provides exact semi-analytical solutions. However, the general approach outlined in this paper is by no means restricted to a certain shape or material composition of either photonic or plasmonic elements. Importantly, unlike ab initio numerical methods, GMT provides insight into underlying physical processes and enables comparing contributions from different scattering and coupling mechanisms within complex photonic, plasmonic and hybrid structures.

### 3. Collective field effects in hybrid optoplasmonic structures

Microcavity WG modes interact with the environment through their evanescent 'tails,' which can extend into the cavity exterior by tens or hundreds of nanometers (depending on the sphere size and index contrast), with the field intensity dropping off exponentially away from the sphere surface. Interaction of the WG mode with the noble metal nanoparticles immersed into its evanescent mode tail can lead to the excitation of localized SP resonances on the nanoparticles, resulting in the cascaded field enhancement in hybrid optoplasmonic structures [20, 21, 23, 39-41]. This is illustrated in Fig. 2, where the electric field intensity spectrum in a hybrid optoplasmonic structure composed of an Au nanoparticle dimer and a dielectric microcavity (Fig. 2(a)) is compared to those of an identical isolated dimer and of a stand-alone microsphere (Fig. 2(b)). Experimentally obtained Au refractive index values from Johnson and Christy are used in all the simulations [42]. The additional microcavity-induced field enhancement of the dimer LSP field is driven by the long dephasing time of the high-Q microcavity mode, which enables efficient accumulation and recirculation of energy from external sources, albeit at the price of narrowed frequency bandwidth. Cascaded nanoscale hot-spot intensity enhancement and resonant manipulation of radiative rates of embedded emitters in hybrid optoplasmonic structures have recently been predicted theoretically [20, 23, 24, 41, 43] and demonstrated experimentally [21, 40, 44-50]. These effects provide far-reaching opportunities for developing optical nanotraps and sensors with simultaneously high sensitivity and high spectral resolution [21, 40] as well as functional elements for future quantum information optoplasmonic networks [20].

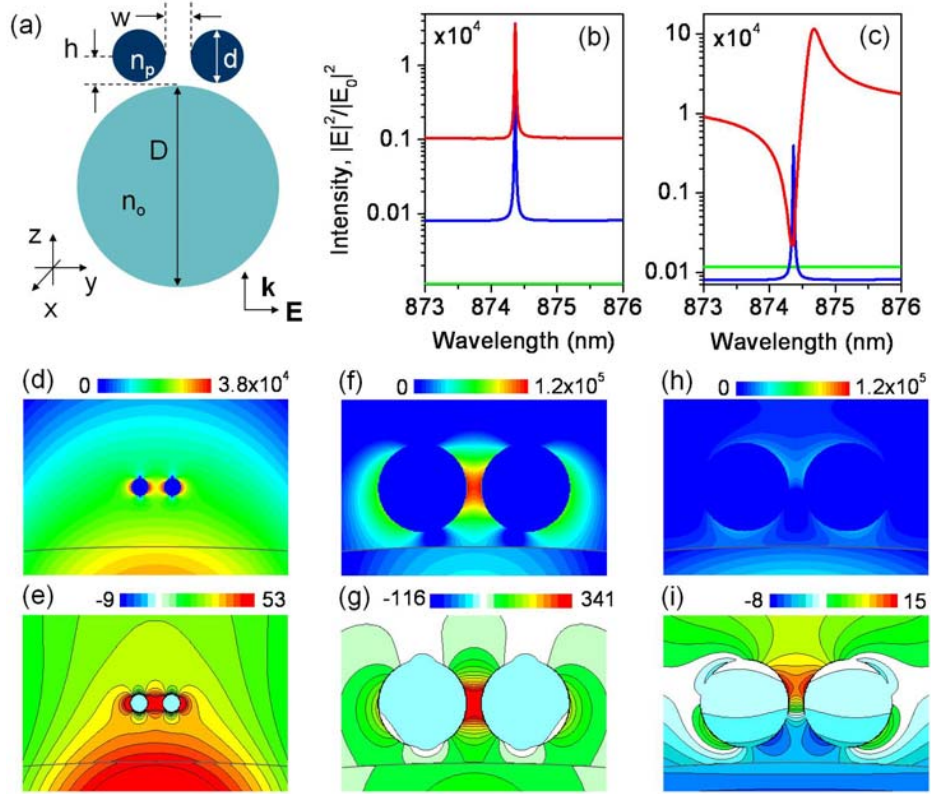

Fig. 2. Formation of photonic-plasmonic modes in hybrid optoplasmonic structures. (a) A schematic of an optoplasmonic structure composed of an Au dimer-gap nanoantenna coupled to a dielectric microsphere. (b,c) Frequency spectra (red) of the intensity enhancement in the dimer gap around $TE_{27,1}$ resonance of the sphere for D=5.6μm, h=100nm, $n_o$=1.59, w=25nm, and d=30nm (b) or d=150nm (c). The corresponding spectra for the isolated dimer (green) and the isolated microsphere (blue) are shown for comparison. (d,e) Spatial maps of $|E|^2$ (d) and $Re(E_y)$ (e) at the wavelength of the resonant peak observed in (b). (f-i) Spatial maps of $|E|^2$ (f,h) and $Re(E_y)$ (g,i) at the wavelengths of the resonant peak (f,g) and dip (h,i) observed in (c).

However, a complex local phase landscape formed around resonantly-excited microcavities offers a high level of control over the delicate sub-wavelength structure of the associated optical field, which goes far beyond the realization of cascaded photonic-plasmonic field enhancement. In particular, complex structured optical spectra (in both, far- and near-field) can be tailored by properly designing individual constituents of optoplasmonic structures as well as by tuning their mutual electromagnetic coupling. For example, the simple model optoplasmonic structure shown in Fig. 2(a) can be engineered to feature Fano-type spectral features rather than Lorentzian-type resonant peaks as in Fig. 2(b). In the simplest case, this can be achieved by increasing the size of the Au nanoparticles and thereby making additional 'dark' hybridized LSP eigenmodes of the dimer available for coupling to the WG mode of the microcavity [51]. A typical sharp Fano-resonance lineshape appearing in the spectrum of the optoplasmonic structure at the wavelength of the microsphere WG-mode resonance is shown in Fig 2(c). The spatial maps of the electric field intensity around the Au nanodimer at the wavelengths corresponding to the peak intensities observed in Figs. 2(b,c) are shown in Figs. 2(d,f). In addition, we plot spatial maps of the real part of the principal electric field component $\text{Re}\{E_y\}$ in Figs. 2(e,g) to reveal the parity of the hybridized modes.

It can be seen that the formation of the intensity peaks in both Fig. 2(b) and (c) reflect coupling of the microcavity WG mode to a bright hybridized bonding LSP mode in the dimer. This bonding LSP mode is characterized by a positive parity of dipole moments oriented along the dimer axis and by a symmetric electric field distribution [51]. The dip in the intensity spectrum of the optoplasmonic structure with larger-size Au nanoparticles (observed in Fig. 2(c)) is the manifestation of the WG mode coupling to a hybridized SP mode with a field distribution that is asymmetrical with the respect to the dimer axis and symmetrical in the perpendicular direction (Figs. 2(h,i)).

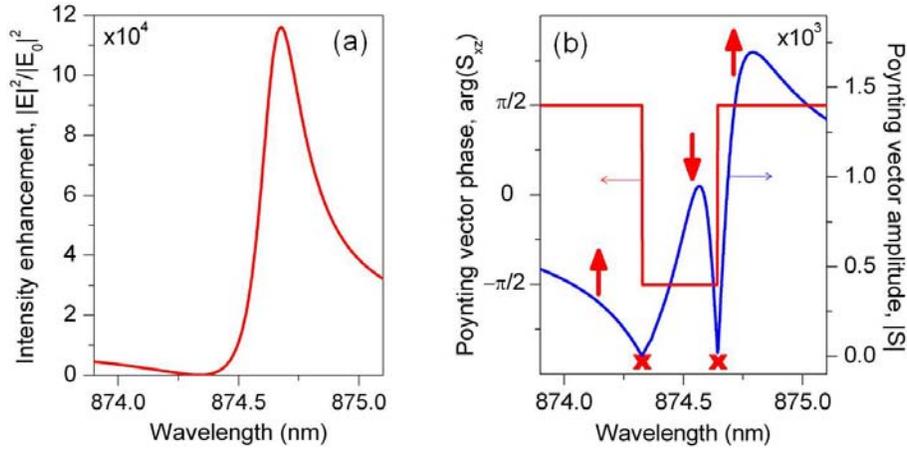

Fig. 3. Manipulation of optical powerflow through the nanoantenna gap with vortex nanogates. (a,b) Comparison of the intensity enhancement spectrum in the gap of microsphere-coupled Au dimer (a) with those of the Poynting vector amplitude and phase at the gap center (b) (D=5.6μm, h=100nm, w=25nm, d=150nm, $n_o$=1.59). Crosses mark the wavelengths at which there is no optical powerflow through the gap ('gate closed'), and arrows indicate the direction of the powerflow in the corresponding frequency range ('gate open Up' or 'gate open Down').

The spectrum in Fig. 2(c) clearly demonstrates that hybrid optoplasmonic structures can provide switching capabilities with ultra-large modulation depth within ultra-narrow frequency bands. This paves the road to the development of all-optical memory nanoelements and on-chip optoplasmonic switching architectures with high-contrast logic levels. Furthermore, a dense spectrum of high-Q modes in microcavities makes possible combining high wavelength selectivity with multi-channel operation and multiplexing/demultiplexing capabilities [20]. The high Q-factors of the microcavity modes (resulting in their narrow linewidths) help to drastically reduce the value of the smallest refractive index change that is required to perform switching operation [34, 52] and thus can overcome the intrinsic material limitation of noble metals in active plasmonic nanostructures. Depending on the microcavity material, the dynamic refractive index change can be achieved all-optically (by photo-injection) [52], electro-optically [53, 54] or via temperature-induced changes [55].

## 4. Optoplasmonic vortex nanogates

To gain deeper insight into the dynamic picture of modes coupling and hybridization in the optoplasmonic structure in Fig. 2(a) with the change of external control parameters, we traced the evolution of the optical powerflow through the gap of the Au nanodimer as a function of the excitation wavelength. In Fig. 3, we compare the intensity spectrum in the dimer gap (Fig.

3(a)) with the corresponding spectra of the amplitude and phase of the Poynting vector at the gap center (Fig. 3(b)). The amplitude of the three-dimensional (3D) real-valued Poynting vector $\mathbf{S} = \{S_x, S_y, S_z\}$ at a given point of space defines the local value of the optical power density. In turn, the Poynting vector phase can be defined in each coordinate plane as $Arg(S_{ij}) = \arctan S_j/S_i$, $i, j = x, y, z$, and characterizes the direction of the optical powerflow through this point. In particular, $Arg(S_{xz}) = \pm \pi/2$ (corresponding to $Arg(S_{yz}) = \pm \pi/2$) indicates either upward or downward energy flow through the dimer gap, respectively.

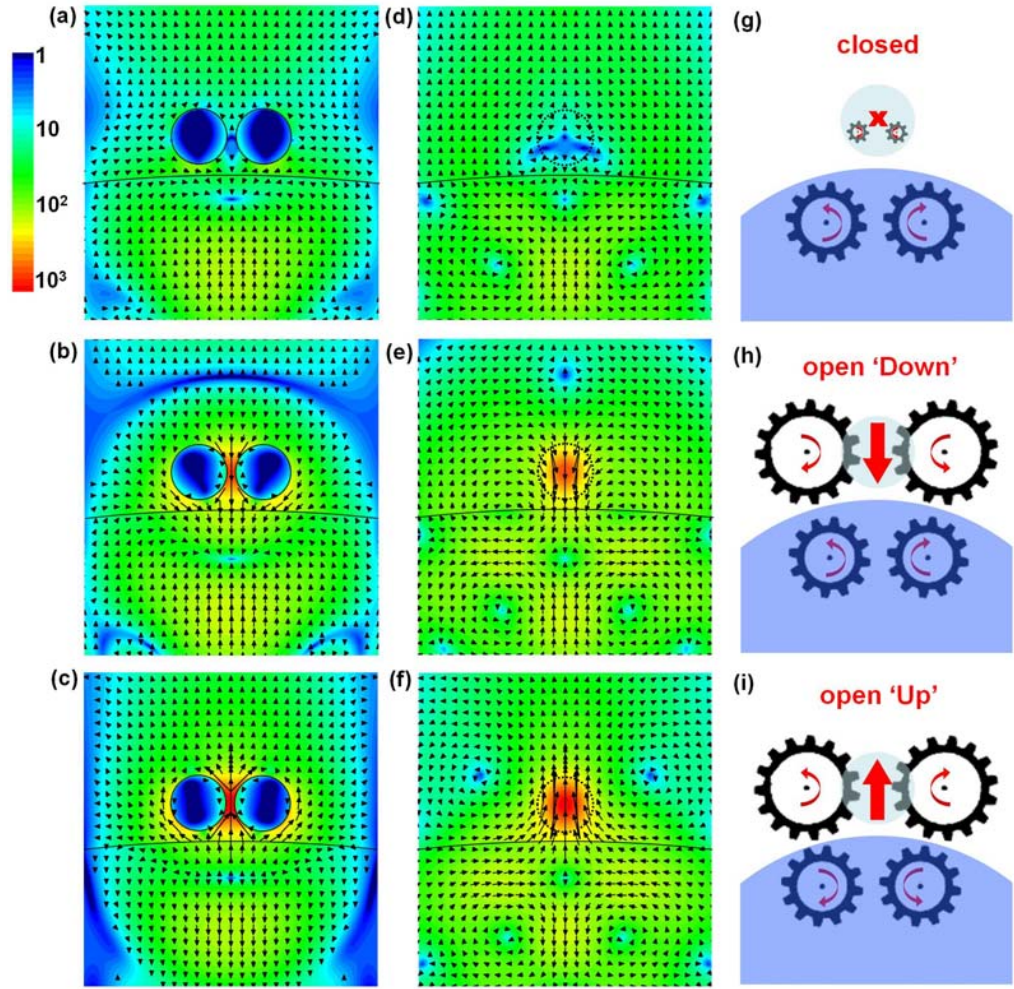

Fig. 4. Operation of the optoplasmonic vortex nanogate. (a-f) Single-frame excerpts from movies of the Poynting vector intensity |S| maps and the optical power flow through the nanoantenna gap at the frequencies around the photonic-plasmonic Fano resonance shown in Fig. 3. The arrows point into the direction of the local powerflow, and the length of each arrow is proportional to the local value of the Poynting vector amplitude. Spatial maps are shown in the y-z plane at x=0 (a-c) and in the x-z plane at y=0 (Media 3, d-f), respectively. Solid circles indicate the boundaries of the Au nanospheres, and the dotted circles are the projections of the nanospheres on the plane cutting through the center of the dimer gap. (g-i) Schematics of the vortex-operated nanogates in the 'closed', 'open Down' and 'open Up' positions.

Comparison of the spectra in Figs. 3(a) and (b) reveals a complex picture of alternating light flow through the dimer gap with the change of the wavelength. The direction of the powerflow reverses from the forward-moving at the frequencies both above and below the resonance to the backward-moving in a narrow range around the resonance wavelength. Essentially, the optoplasmonic structure driven across its hybridized photonic-plasmonic resonance operates as a photonic nanogate, which directs the optical powerflow through the nanosized channel (dimer gap). The gate can be open either 'Up' or 'Down', corresponding to either forward or backward powerflow, or 'Closed' at select wavelengths where the Poynting vector amplitude vanishes (resulting in no powerflow through the gap). It should be noted that the symmetrical coupled-dipole LSP resonance of the isolated nanoparticle dimer is characterized by the enhanced forward powerflow at the resonance frequency.

To better understand the physical processes underlying the operation of the optoplasmonic nanogate, in Fig. 4 we plot the spatial maps of the Poynting vector amplitude and powerflow through the nanodimer at select frequencies around the resonance. Here, each arrow points into the direction of the local powerflow, and its length is proportional to the local value of the power density. The complex 3D structure of the powerflow can be revealed by considering the 2D flow distributions in two perpendicular planes: one cutting along the dimer axis (y-z plane) and the other cutting through the center of the dimer gap (x-z plane) as shown in Figs. 4(a-c) and Figs. 4(d-f), respectively. The powerflow distributions in the y-z plane feature enhanced(suppressed) flow through the dimer gap corresponding to the open (closed) nanogate state. The most revealing, however, is the evolution of the powerflow in the x-z plane, which features formation, evolution and disappearance of local areas of circulating powerflow – optical vortices [33, 56, 57] (Media 3, Figs. 3(d-f)).

The physical origin of the optical vortex is the simple fact that the optical energy flows in the direction of the phase change. Therefore, phase singularities (occurring at points of zero field intensity) are always accompanied by the circulation of the optical energy [33, 56, 57]. Free-space optical vortices occur in the interference field arising from the superposition of three or more partial waves, and can be controllably created by using "forked" holograms, lenses, spiral phase plates, and spatial light modulators. Electromagnetic fields diffracted by photonic and plasmonic nanostructures can also feature optical vortices resulting from the superposition of incident and refracted/reflected waves [58, 59]. In the considered optoplasmonic nanogate, optical vortices form around the phase singularities emerging in the interference field around the microcavity excited at the wavelength of its WG mode (shown in Fig. 1(d)). In fact, the well-known $\exp\{im\varphi\}$ phase-dependence of WG modes [60] is a signature of the presence of phase singularities in their on-resonance electromagnetic field. In this context, the azimuthal mode index $m$ gives the number of $2\pi$ phase cycles around the optical vortex formed around the microcavity center. However, high-strength vortices – i.e. those with $m > 1$ – are unstable, and typically break up into $|m|$ vortices of strength $\text{sign}(m)$ upon perturbation by an external field featuring no vortices (e.g., a constant field or a plane wave).

Media 3 shows that under excitation by a plane wave with the wavelength approaching that of the hybridized photonic-plasmonic resonance, pairs of coupled counter-rotating optical vortices are formed inside the microcavity. This results in the enhanced backward powerflow just below the microcavity surface and reduces the E-field intensity generated in the gap of the Au nanoparticle dimer located above the surface (observed in Fig. 3(a)). Tuning of the excitation wavelength across the resonance leads to the nucleation of a pair of vortices of the opposite rotation direction in the Au dimer gap, which completely block the powerflow through the gap (Fig. 4(a,d)). With the increase of the wavelength, these vortices drift apart from each other and closer to the microcavity surface, and their combined effect yields enhanced backward powerflow through the dimer gap (Fig. 4(b,e)). Note that the nucleation and annihilation of vortices is accompanied by the formation and disappearance of the powerflow saddle points, in accordance with the topological charge conservation principle [58, 61]. In particular, a saddle point located just below the Au nanodimer is clearly visible in

Figs. 4(a-f). Another saddle point accompanies the emergence of the vortex pair in the dimer gap. This point drifts up and down above the dimer, and its position defines the extent of the area with the enhanced backward powerflow (Fig. 4(b,e)). If the wavelength is further increased, the vortices migrate from the microcavity surface into the microcavity material, which once again reverses the powerflow direction through the dimer gap. At the same time, oppositely-rotating vortices form in the outer evanescent-tail area, yielding the enhanced forward powerflow through the gap as shown in Fig. 4(c,f). This enhanced forward powerflow drives the intensity enhancement at the resonant peak observed in Fig. 3(a). When the excitation wavelength moves away from the resonance, the phase singularities approach and annihilate, resulting in a reduced powerflow through the gap, which leads to a reduction of the near-field intensity in the nanodimer gap.

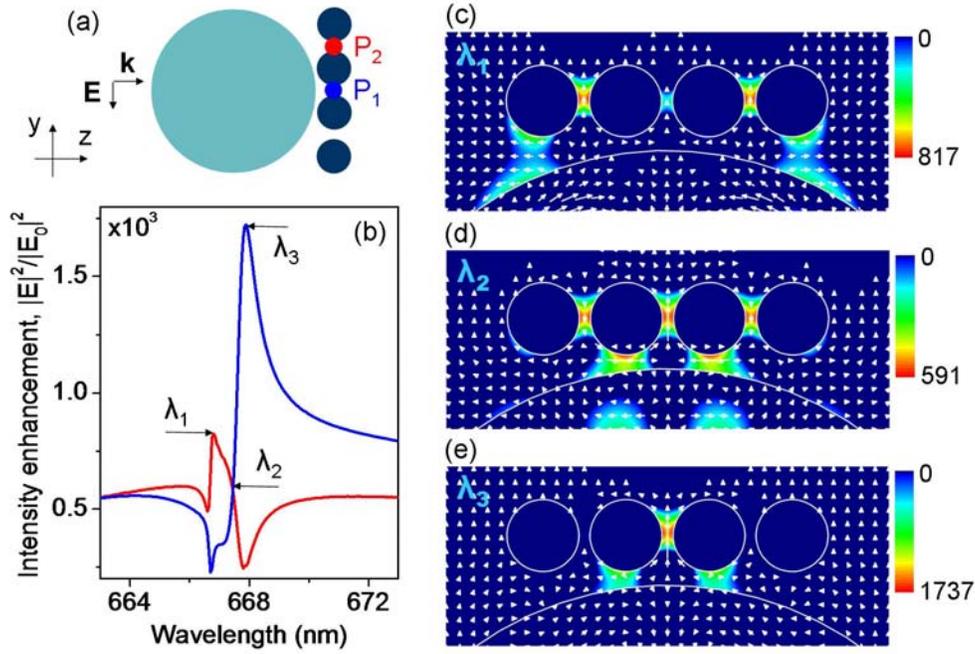

Fig. 5. Phase-operated nanoscale field intensity switching in the plasmonic nanoparticle chain. (a) A schematic of an optoplasmonic structure composed of a linear chain of Au nanoparticles coupled to a dielectric microsphere (D=1.2μm, h=90nm, w=20nm, d=130nm, $n_o$=2.4). The points where the field intensity is monitored are marked as $P_1$ and $P_2$. (b) The near-field intensity spectra evaluated at $P_1$ (blue) and $P_2$ (red). The three select wavelength $\lambda_1$=666.74nm, $\lambda_2$=667.45nm, and $\lambda_3$=667.87nm mark the spectral points where the intensities at $P_1$ and $P_2$ are either equal ($\lambda_2$) or one of them reaches its peak value ($\lambda_1$, $\lambda_3$). (c-e) Single-frame excerpts from the movie (Media 4) showing the evolution of the electric field intensity ($|E|^2/|E_0|^2$) distribution and the optical power flow through the nanoparticle chain in the y-z plane at x=0 at $\lambda_1$, $\lambda_2$ and $\lambda_3$, respectively. The arrows point in the direction of the local powerflow, and the length of each arrow is proportional to the local value of the Poynting vector amplitude.

Various configurations of coupled optical vortices appearing in Figs. 4(d-f) and Media 3 resemble a complex switchable gearbox composed of multiple 'vortex nanogears', which can be dynamically rearranged by tuning the control parameters such as wavelength. The nanogears arrangement drives the local powerflow through the optoplasmonic nanogates, which can be reversibly switched into either 'Open Up/Down' or 'Closed' positions. The coupled vortex nanogears configurations corresponding to different regimes of the nanogate operation are schematically visualized in Figs. 4(g-i).

## 5. Adaptive spatial light control in optoplasmonic networks

The demonstrated ability of optoplasmonic elements to combine efficient nanofocusing of the optical energy with the strategies developed in this article to dynamically mold and re-direct the energy flow on the nanoscale can be used to develop more complex reconfigurable elements and networks. An example of such a reconfigurable optoplasmonic element is shown in Fig. 5 (a) and represents a linear chain of Au nanospheres coupled to a microcavity. Unlike the simplest structure shown in Fig. 2, this configuration features more than one focal point (dimer gap). The excitation of the isolated linear nanoparticle chain with the external plane wave results in the formation of hybridized LSP modes, with the most prominent low-Q bonding symmetrical mode featuring elevated field intensity in all the gaps along the chain (see e.g., [62]). The presence of the high-Q photonic element can be used to resonantly manipulate the powerflow through the nanoparticle chain, and, thus, to dynamically re-configure the spatial intensity distribution of the nanofocused light.

This effect is demonstrated in Fig.5(b), which shows the wavelength spectra of the near-field intensity calculated at two spatial locations: in the central and side inter-particle gaps, respectively (labeled as $P_1$ and $P_2$ in Fig. 5(a)). It can be seen that at the wavelength corresponding to the excitation of the $TM_{10,1}$ WG-mode in the microsphere, the local field intensity plots feature high-amplitude oscillations. Comparison of the spectra at $P_1$ and $P_2$ reveals a frequency-dependent spatial intensity switching between the central and the neighboring interparticle gaps. The electric field intensity distributions at the frequencies corresponding to the light focusing in the side gaps, even distribution between all the gaps, and focusing in the central gap are shown in Figs. 5(c), (d), and (e), respectively. The intensity patterns in Figs. 5(c-e) are overlapped with the corresponding optical powerflow maps (white arrows), which show that the intensity switching is driven by phase-controlled re-configuration of the local powerflow through the nanostructure.

## 6. Conclusions

We proposed and demonstrated novel optoplasmonic structures that exploit the rich phase landscape of the near-field of resonantly-excited optical microcavities to controllably manipulate sub-wavelength spatial light distributions in nanoscale structures. Our results demonstrate that the nanoscale powerflow through plasmonic structures can be directed and reversibly switched with *chip-integrated* high-Q photonic elements (microcavities). The possibility of *local addressing* of individual microcavities (optically, electro-optically or thermo-optically) in a dynamic fashion is the advantage of the proposed mechanism of the phase-operated intensity switching over previously explored strategies based of using external phase- and amplitude-modulated pulses and beams. Although we explored this approach in a few selected configurations of optoplasmonic elements in this article, the proposed strategy is very general and can be applied to design extended optoplasmonic networks of arbitrary morphology that incorporate various types of microcavities and plasmonic nanostructures. Our observations pave the road to the development of dynamically-tunable and switchable vortex-operated plasmonic nanocircuits for optical information processing and ultrasensitive biosensing.

## 7. Acknowledgments

The work was partially supported by the National Institutes of Health through grant 5R01CA138509-03, the National Science Foundation through grants CBET-0853798 and CBET-0953121 (BMR) and by the EU COST Action MP0702 "Towards functional sub-wavelength photonic structures" (SVB). SVB thanks Anton Desyatnikov for stimulating discussions.